\font\smc=cmcsc10 scaled 913
\magnification= \magstep1
\baselineskip=15pt
\hsize=6.truein
\vsize=9.truein
%
\newskip\tableauskip \tableauskip=1em                                           
\newskip\tableauht \tableauht=9.5pt                                             
\newskip\tableaudp \tableaudp=4.5pt                                             

\def\tableau{\begingroup\vcenter\bgroup\offinterlineskip                        
                 \tabskip=0pt                                                   
       \def\trait{\noalign{\hrule}}                                             
       \def\tableausk{\hskip\tableauskip}                                       
       \def\hauteur{\hbox{\vrule height \tableauht                              
               depth \tableaudp width 0pt}}                                     
       \halign\bgroup\vrule##&                                                  
               \hauteur\tableausk$\hfil##\hfil$\tableausk\vrule&&               
               \hauteur\tableausk$\hfil##\hfil$\tableausk\vrule\cr}             
\def\fintableau{\crcr\egroup\egroup\endgroup}                                   
\newcount\numpub
\def\aut#1*{\advance \numpub by 1
            \smallskip
            \hangindent 1.5 truecm
            \noindent \hskip 1 truecm
            \llap{$[$\the\numpub$]$\hskip.35em}
            {\smc#1}}
\def\suitaut#1*{\par
                \hangindent 1.5 truecm
                \noindent \hskip 1 truecm
                {\smc#1}}
\def\livre#1*{\enskip ---\enskip #1\ignorespaces}
\def\editeur#1*{\enskip ---\enskip {\it #1}}
%
%
\def\tita#1{\goodbreak\vskip .6truecm \noindent
             {\bf #1}\nobreak\vskip .4truecm}
%
%
\def\titc#1{\goodbreak\medskip\noindent
             {\it #1}\nobreak\smallskip}

\def\Videbas{\vrule height 0pt width 0pt depth 15pt}
\def\videhaut{\vrule height 12.5pt width 0pt depth 0pt}
\def\videbas{\vrule height 0pt width 0pt depth 7.5pt}

{\nopagenumbers

\null\vskip 1truecm
\centerline{\bf About the determination of critical exponents related}
\medskip
\centerline{\bf to possible phase transitions in nuclear fragmentation}
\vskip 2truecm

\centerline{B. Elattari}
\smallskip
\centerline{Max-Planck Institut f\"ur Kernphysik, Postfach 103980}
\smallskip
\centerline{69029 Heidelberg, Germany}
\vskip 1truecm
\centerline{J. Richert}
\smallskip
\centerline{Laboratoire de Physique Th\'eorique, 3 rue de l'Universit\'e,}
\smallskip
\centerline{67084 Strasbourg Cedex, France}
\vskip 1truecm
\centerline{P. Wagner}
\smallskip
\centerline{Institut de Recherches Subatomiques, BP 28}
\smallskip
\centerline{67037 Strasbourg Cedex 2, France}
\vskip 2.5truecm
\noindent
{\bf Abstract}. We introduce a method based on the finite size scaling
assumption which allows to determine numerically the critical point and 
critical exponents related to observables in an infinite system starting from the
knowledge of the observables in finite systems. We apply the method to bond
percolation in 2 dimensions and compare the results obtained when the bond
probability $p$ or the fragment multiplicity $m$ are chosen as the relevant
parameter. 

\vfill\eject}
  
\pageno=1
  
\tita{1. Introduction}
  
Phase transitions are phenomena which are related to infinite systems. Hence
the quantities which characterize these phenomena must be related to infinite
systems. In fields like solid state physics this can be more or less easily
realized since the systems which are considered can most of the time be
assimilated to infinite systems. But this is not the case in nuclear physics
where nuclei contain at most a few hundred constituents.
\medskip
Among the most striking features which are observed in the study of nuclear 
fragmentation emerges the fact that there are signs for the existence of a
phase transition. This phenomenon is commonly under scrutiny nowadays, both
experimentally and theoretically. There are indications for the existence of
a first order transition but its existence is not clearly established up to
now [1-4]. On the other hand, charge distributions of fragments obtained
in peripheral collisions show universal properties [5] which are
conspicuously compatible with bond percolation [6] and phase space models
which describe excited and disordered systems [7,8]. If one relies on a
percolation-type interpretation of these results, it is tempting to conclude
that one is confronted with the existence of a second order phase transition
whose characteristic features are smoothed by finite size effects.
\medskip
The fascinating perspective that it might be possible to unravel the
existence of a phase transition in nuclear matter motivated the attempt to
extract more precise information about this kind of phenomenon. Such information
is given by the numerical values of critical exponents which allow to fix the
class to which the observed phase transition belongs. The most important
exponents are $\nu$ which governs the correlation length $\xi$, $\sigma$
which characterizes the behaviour of $m_2/m_1$ ($m_k$ is the moment of order
$k$ of charge or mass distributions), $\beta$ which fixes the behaviour of
$m_1$ close to the critical point, $\tau$ which gives the slope of the power
law distribution at the critical point and $\gamma$ which governs the strength
of the singularity of $m_2$ at this point. These quantities are connected by
universal relations [16].
\medskip
The determination of critical exponents raises however two intimately related
points. The first one concerns the obtention of these quantities from observables
which characterize finite systems and hence do not show any singularity as
they do for infinite systems. This is a conceptual problem which, to our
opinion, is fundamental. The second point which is directly linked to it
concerns the direct extraction of exponents from finite size model results or
experimental data necessarily obtained from systems of finite size $L$,
especially in the case of nuclei. Any such attempt leads inevitably to a 
size-dependence of any ``exponent" say $\alpha$, i.e. $\alpha=\alpha\,(L)$.
As a consequence, it gets difficult if not impossible to compare these quantities
to corresponding reference exponents characterizing well-known systems, like
the liquid-gas, percolation or magnetic systems, which have been worked out 
in the infinite limit and fix the universality class to which the considered
system belongs. The possibility to extract critical exponents from finite
size systems has been attempted in refs. [9-11] and was followed by 
controversial debates [12,13].
\medskip
In the sequel we aim to present and use a new numerical approach which has
been worked out recently in order to analyze finite size calculations in the
2D and 3D Ising models [14]. The method relies on the finite size scaling (FSS)
assumption [15].
  
\tita{2. The method}
  
We sketch here the essential steps which lead to the determination of the 
critical point and critical exponents for infinite systems starting from
finite systems of different sizes $L$ in spaces of any physical dimension.
\medskip
Consider an observable $O_L(\Delta t)$ characterizing a finite system of linear size 
$L$ where $\Delta t$ measures the distance of the variable $t$ to its value
at the critical point $t_c$, $\Delta t = (t_c-t)/t_c$, $\Delta t >0$. Define
$x\,(L,t)\equiv \xi_L(t)/L$, where $\xi_L(t)$ is the correlation length in
the system of size $L$ for a fixed value $t$ of the relevant variable. Then the FSS
assumption states that
$$O_L(\Delta t) = O\,(\Delta t)\cdot Q_O(x\,(L,t))\eqno(1)$$
  
\noindent
where $Q_O(x\,(L,t))$ is a so called scaling function and $O\,(\Delta t)$
fixes the value of the observable in the corresponding infinite system. The
remarkable property of $Q_O(x)$ is the fact that it is universal, which means
here that it is independent of $t$ for different values of $x$ in the range
$[0, t_c]$ of the variable $t$.
\medskip
It comes out that quantities like $\xi_L(\Delta t)$ and other observables
$O_L(\Delta t)$ converge to a finite value with increasing $L$. Hence, if
$O_L(\Delta t)$ is known for different values of $L$ and if its asymptotic
value with increasing $L$, $O\,(\Delta t)$, can be numerically reached the 
scaling function $Q_O(x\,(L,t))$ can be obtained through (l). Once this
quantity is known for different values of $x\equiv \xi_L(t)/L$, it can be
used to get 
$$O\,(\Delta t) = O_L(\Delta t)/Q_O(x) \sim \Delta t^{-\alpha} \sim
(t_c-t)^{-\alpha}\eqno(2)$$
  
\noindent
where $t_c$ is the critical point and $\alpha$ the corresponding exponent.
\vfill\eject
  
In practice, once $Q_O(x)$ is numerically known it can be easily parametrized
in terms of polynomials in $x$ or some other function of $x$ [14]. The
knowledge of $Q_O$ and $O_L(\Delta t)$ gives $O\,(\Delta t)$ through (2), the
numerical result can be fitted to an expression like $(t_c-t)^{-\alpha}
$ through
which $t_c$ and $\alpha$ can be determined. The accuracy of the fits can be
controlled by means of $\chi^2$ tests.
  
\tita{3. Application to 2D percolation}
  
We have checked the applicability of the method presented above on the 2D
bond percolation problem on a square lattice for which the numerical values
of the critical exponents are well known [16]. In the sequel we restrict
ourselves to the explicit calculation of
$\nu$, $\gamma$, $\sigma$ and the critical point.
We try to obtain these quantities through the determination of $\xi_L(\Delta t)$, 
$m_{2L} (\Delta t)/L^2$ and $m_{2L} (\Delta t)/m_{1L} (\Delta t)$ where $L$ is
the linear size of the 2D lattice and $t$ stands either for $p$ (bond probability)
or $m$ (cluster multiplicity).
\medskip
The choice of $m_{2L} (\Delta t)/L^2$ deserves a comment. In practice we are
interested in $m_{2L} (\Delta t)$, but this quantity does not converge to any
finite value when $L$ goes to infinity, whatever $t$ (extensive quantity). We
found no rigorous mathematical proof for the convergence of the aforementioned
quantity to a constant asymptotic value with increasing $L$. It is however
trivial to see that $m_{2L}/L^2 = 1$ when $t=p=0$. It is equal to 0 when 
$t=p=1$ (the heaviest cluster is not taken into account). Furthermore, numerical
tests indicate that this quantity may indeed reach an asymptotic value, although 
this value is reached for larger and larger values of $L$ when
$t\,(p,m)$ gets closer and closer to $t_c$. We discuss this point in the
sequel. If this convergence property is fulfilled, one faces a second point
which is the question whether the exponent and location of the critical point
are the same for $m_{2L}$ and $m_{2L}/L^2$ when $L$ is infinite. One can
convince oneself that the corresponding scaling functions $Q$ are not the
same. The tests presented below confirm that $t_c$ and the exponent are in
numerical agreement with the known values [17]. Finally, it is easy to
show that $m_{1L}/L^2$ tends to 1 with increasing $L$. Hence $m_{2L}/m_{1L}$
reaches a constant limit for very large systems and allows to extract in
principle the exponent ($\sigma$ or $\tilde\sigma$, see below).
\medskip
Calculations and the
results which will be shown correspond to systems without periodic boundary
conditions. Comments about calculations made for systems with periodic boundary 
conditions will be made below.
\medskip      
\titc{3a. Calculations with the bond probability p}
\smallskip  
We have determined the correlation length $\xi_L(\Delta p)$, the normalized
second moment $m_{2L}(\Delta p)/L^2$ and the ratio $m_{2L}(\Delta p)/m_{1L}(\Delta p)$
for $L$ up to 1$\,$000 and different values of $p$ in the interval [0.40, 
0.48] in steps of $\Delta p=0.01$. For each value of $p$ we generated $10^4$
events. 
The critical value is $p_c=0.5$.
Fig. 1 shows the behaviour of the obser\-vables. As it can be seen, these quantities
tend to reach a constant value. One may notice that this saturation effect is
better realized for smaller values of $p$ and more effective for $\xi_L$ than
for $m_{2L}$. As it can be seen on the figure, it is not effectively reached
for the highest value of $p$. The consequences of these observations will
be seen below.
\bigskip
We retain the asymptotic values reached by the observables as those which
correspond to the infinite system and use them in order to get the universal
scaling functions $Q_\xi$, $Q_{m_{2L}/L^2}$ and $Q_{m_{2L}/m_{1L}}$ through
(1). Fig. 2 shows examples of these quantities obtained for different values of $p$
and represented as functions of $x = \xi_L/L$. Universality requires that all
contributions from different $p's$ lie on each other and asymptotic saturation
requires that the $Q's$ reach a constant value (equal to 1) for $x$ close
to 0. The universality functions are fit to polynomial in $x$, see eq. (14) in
ref. [14]. Finally these quantities are used in order to determine 
$p_c, \nu, \gamma$ and $\sigma$ by means of a fit procedure to the quantities
relevant to the infinite system, i.e.
   
$$\eqalignno{
\xi_\infty~ &\simeq~ (p_c-p)^{-\nu}\cr
m_{2\infty}~ &\simeq~ (p_c-p)^{-\gamma}\cr
\noalign{\hbox{and}}
m_{2\infty}/m_{1\infty}~ &=~ (p_c-p)^{-1/\sigma}\cr}$$
   
The quality of the fits is estimated by means of $\chi^2$ analyses. Fig. 3
shows a typical example of the calculations. In practice we worked with many
different intervals of $p$ values in the range [0.40, 0.48] over which we
averaged. We retained those results which correspond to the smallest $\chi^2$
values, with a confidence level of 95\%.
\medskip
Results are shown in Table 1 and Fig. 4. The exponents are calculated from
systems of
different sizes. They are in an overall good agreement with the exact values
[17]. One notices however that $\nu, \gamma$
and $1/\sigma$ decreases with increasing $L$, $\nu$ gets closer to the exact
value while $\gamma$ and $1/\sigma$ get away from it. Introducing periodic
boundary conditions leads to values which remain very close to those which
do not take them into account. In practice, the two types of calculations differ
essentially in the way asymptotic saturation of the observables is reached,
the rise being steeper when periodic boundary conditions are taken into account.
\medskip
\titc{3b. Analysis and discussion of the FFS method}
\smallskip 
As mentionned above,
we have tested the present method by choosing different intervals in $p$
lying more or less close to $p_c$. If the interval is taken for, say between
0.30 and 0.40, the exponents are robust with respect to the value of $L$ from
which they are extracted. The exponents are however sizably smaller
than the exact known values [19].
If the interval approaches $p_c$, they come closer to the exact value but they
are less robust with respect to $L$, as already mentionned. 
\medskip
These ascertainments have a common origin. In principle, the universal functions
$Q_O$ are obtained from the asymptotic constant limit of $O$. This limit with
increasing $L$ is easily reached when $p$ is small. If however $p$ comes
close to $p_c$, higher values of $L$ are needed in order to reach this limit. It comes
out that this limit is more easily reached for $\xi_L$ than for $m_{2L}/L^2$
and $m_{2L}/m_{1L}$ This reflects on the behaviour of $Q_\xi$, $Q_{m2L/L^2}$
and $Q_{m2L/m1L}$. Indeed, one can observe that for values of $L$ up to 1$\,$000, 
$Q_\xi$ reaches a plateau for the smallest values of $x=L/\xi_L$, whereas
one can hardly see this effect for $Q_{m2L/L^2}$ and $Q_{m2L/m1L}$. This means in
practice that one needs to determine the considered observables for values of
$L$ which are sizably larger than $L=1\,$000. We did not work these cases since it is
not our aim to work the exponents for 2D percolation which are well known, 
but rather test a method and its efficiency.
\medskip 
It is possible to find an explanation for the difference in the asymptotic
behaviour of the considered observables. It is easy to observe that
the rate of convergence with $L$ decreases with increasing values of the exponents.
This fact can also be observed in the determination of exponents in ref. [14].
Indeed $\gamma$ and $1/\sigma$ are much larger quantities than $\nu$. Larger
exponents correspond to steeper increase of the corresponding observable,
hence a larger sensitivity to the precise numerical determination of this
observable. One may remark that the exponent $\gamma$ is sizably smaller
($\simeq 1.8$) in 3D bond percolation systems. We conjecture that one may need
smaller values of $L$ in order to reach the asymptotic regime.
\vfill\eject
\titc{3c. Calculations with the multiplicity m}
\smallskip
A confrontation of a percolation model determination of observables like
$m_2$ and $m_2/m_1$ with experiment [10] is possible if one uses the
fragment and particle reduced multiplicity $m$ [20]
(multiplicity divided by $L^2$). It is numerically easy to
relate $p$ to an average value of $m$, $<m>$. It is then tempting to
work out the behaviour of the quoted observables $\xi_L$, $m_{2L}/L^2$ and
$m_{2L}/m_{1L}$ and to confront the exponents $\tilde\nu$, $\tilde\gamma$ and
$\tilde\sigma$ with $\nu$, $\gamma$ and $\sigma$ obtained with the bond
probability $p$.
\medskip
We repeated the preceeding calculations with $m$, following the same procedure
as described in section 3a. For a given $m$ the events were selected out of
a uniform distribution in $p$. For each value of $m$ we generated 
$2\,500$~events.
The same remarks as those presented above are valid
in the present case. However, as it can be seen in Table 1, the corresponding
exponents $\tilde\nu$, $\tilde\gamma$ and $\tilde\sigma$ are no longer 
numerically compatible with $\nu$, $\gamma$ and $\sigma$. In fact, $\tilde\nu$
and $\tilde\gamma$ are systematically smaller than $\nu$ and $\gamma$ whereas
$\tilde\sigma$ is larger than $\sigma$. This result contradicts the findings
presented in ref. [12]. One can convince oneself that the discrepancy
cannot be attributed to the convergence trouble quoted above, this discrepancy
remains for $\tilde\nu$ corresponding to $\xi_L$, for which the saturation 
problem is much less acute than for $\tilde\gamma$ and $\tilde\sigma$. In fact,
this result is not really surprising for us. Indeed, the relation between $p$ and 
$<m>$ is definitely not linear in the neighbourhood of $p_c$ as it seems
to be for 3D percolation on Fig. 7 of [12].
\medskip
The calculations have been repeated for systems with periodic boundary 
conditions. As before, the values of the exponents lie within the error bars
obtained in the case of open boundaries.
  
\tita{4. Summary, discussion and conclusions}
  
We have used a rather simple method relying on the finite size scaling (FSS)
assumption which has earlier been tested on the 2D and 3D Ising model [14].
The method allows the rigorous determination of critical exponents
characterizing second order phase transitions in genuine infinite systems.
\medskip
The method is numerically simple, in principle robust, and should deliver 
exponents with a very good precision. We verified that it is indeed possible
to get values which lie within a few percent of the exact value with a rather modest
amount of work. The central point lies in the fact that one must determine the
asymptotic value of the corresponding observables as cleanly as possible.
This means that it is necessary to determine these observables for larger and
larger systems. The maxi\-mum size corresponding to the asymptotic (infinite size)
regime depends on the observable itself and empirical experience shows that it
depends upon the magnitude of the exponent itself. There exists of course
methods which give much more precise exponents [17,18] but they are
numerically much more sophisticated than the present one.
\medskip
In the present work we have concentrated on the academic example of 2D bond
percolation and seen that the exponents depend on the relevant parameter 
(bond probability $p$ or cluster multiplicity $m$) which is used. One may ask
oneself whether and how the FSS method could be applied to a realistic
physical case such as nuclear fragmentation. As we already said in the
introduction, the determination of critical exponents requires rigorous
information about the infinite system. This shows that the determination of
exponents for nuclear systems may be quite difficult. Indeed nuclei are very
small (number of coexisting nuclei $\simeq 300-400$ in heavy ion collisions),
unlike systems found in condensed matter physics where surface effect can be made
negligible if the considered samples are large enough.
\medskip
The necessary reference to the infinite system also shows that there exists
no direct way to extract exponents from experimental data. One way one may
think about consists of working out a model which reproduces the experimental
observables obtained for fixed finite sizes $L$. One may then extrapolate
the model to $L = \infty$ in the FSS framework and determine the exponents
the way which was described above. If the observables can be determined
experimentally for several sizes (f.i. $L_1$ and $L_2$) and if they are in
agreement with the model then, using the notations of section 2, these
observables must verify the scaling relation
$$O^{\exp}_{L_1} (\Delta t)/O^{\exp}_{L_2} (\Delta t) = Q_O (x\,(L_1, t))/Q_O
(x\,(L_2, t))$$
  
If this is the case one may conclude that the system is critical in the 
infinite limit and the exponents determined in this limit effectively characterize
the criticality of the system.
\medskip
Whether such tests can be performed or not in practice is an open question
which is strongly related to the precision with which relevant observables
can be experimentally determined.
\vfill\eject
\noindent
{\bf Table caption}
\bigskip

\line{\hskip 0.25truein
\vbox{\hsize = 5.5truein
\itemitem{Table 1 :} 
Location of the critical point and exponents calculated for different
linear sizes $L$ of the 2D percolation system using the FSS assumption (1).
The numbers in parentheses correspond to the estimated error on the last
figure of the given value.\hfill\break
The exact values are :
$$p_c = 0.5~;~\gamma = 2.389~;~\nu = 1.333~;~\sigma = 0.396$$}\hfill}
\bigskip   
\tableauskip=10pt
$$\tableau
\trait
&\videhaut\videbas L&p_c&\gamma&\nu&\sigma\cr
\cr
&\hfill 40&0.505(5)&2.43(6)&1.42(5)&0.40(2)~~~~~~~\cr
&\hfill 80&0.503(5)&2.43(4)&1.40(6)&0.40(2)~~~~ {\rm a)}\cr
&\Videbas\hfill 120&0.501(3)&2.34(7)&1.35(3)&0.42(3)~~~~~~~\cr
&\hfill 40&0.505(5)&2.41(8)&1.38(7)&0.41(4)~~~~~~~\cr
&\hfill 80&0.500(2)&2.41(5)&1.39(4)&0.43(3)~~~~ {\rm b)}\cr
&\videbas\hfill 120&0.500(2)&2.28(8)&1.35(4)&0.44(3)~~~~~~~\cr
\trait
&\videhaut\videbas L&m_c&\widetilde\gamma&\widetilde\nu&\widetilde\sigma\cr
\cr
&\hfill 40&0.088(5)&2.08(4)&1.26(7)&0.46(2)~~~~~~~\cr
&\hfill 80&0.097(3)&2.01(7)&1.17(7)&0.49(3)~~~~ {\rm c)}\cr
&\Videbas\hfill 120&0.098(3)&1.98(4)&1.17(3)&0.50(2)~~~~~~~\cr
&\hfill 40&0.100(7)&1.93(7)&1.15(7)&0.51(5)~~~~~~~\cr
&\hfill 80&0.097(3)&2.00(6)&1.14(5)&0.47(3)~~~~ {\rm d)}\cr
&\videbas\hfill 120&0.102(3)&1.97(5)&1.09(5)&0.51(3)~~~~~~~\cr
\trait
\fintableau$$
\bigskip
\itemitem{a)} bond probability $p$ with open boundaries
\medskip
\itemitem{b)} bond probability $p$ with periodic boundaries
\medskip
\itemitem{c)} reduced multiplicities $m$ with open boundaries
\smallskip
\itemitem{d)} reduced multiplicities $m$ with periodic boundaries

\vfill\eject
\noindent
{\bf Figure caption}
\vskip 1truecm

\itemitem{Fig. 1 :} Evolution of $\xi$ and $m_2/L^2$ as a function of the
linear size $L$ for different values of $p$.
\bigskip
  
\itemitem{Fig. 2 :} Behaviour of the universal functions $Q$ related to
$\xi$ and $m_2/L^2$ as a function of $x$ for different values of $p$.
\bigskip
  
\itemitem{Fig. 3 :} Typical fit of the second moment represented as a function
of ($p_c-p$) on logarithmic scale for $L=40$. The dispersion generated 
through the simulations is smaller than the size of the points.
\bigskip
  
\itemitem{Fig. 4 :} Values of the critical exponents from bond probability
calculations extracted from the simulations for different values of $L$. The 
exact known values are indicated by the lines.
\vfill\eject

\noindent
{\bf References}
\bigskip
  
\aut J. Pochodzalla et al.*
\editeur Phys. Rev. Lett.* {\bf 75} (1995) 1040 
\smallskip
    
\aut X. Campi, K. Krivine and E. Plagnol* 
\editeur Phys. Lett.* {\bf B385} (1996) 1
\smallskip
  
\aut Y.G. Ma et al.*
\editeur Phys. Lett.* {\bf B390} (1997) 41
\smallskip
  
\aut G. Imme et al.*
\livre GSI-Preprint 96-31*, to appear in the Proceedings of the
1st Catania Relativistic Ion Studies, {\it Critical Phenomena and Collective
Observables}, Acicastello, May 27-31, 1996
\smallskip
  
\aut A. Sch\"utthauf et al.* 
\livre GSI-Preprint 96-26*
\smallskip
  
\aut Y.M. Zheng, J. Richert and P. Wagner*
\editeur J. Phys.* {\bf G22} (1996) 505
\smallskip
  
\aut B. Elattari, J. Richert, P. Wagner and Y.M. Zheng*
\editeur Phys. Lett.* {\bf B356} (1995) 181
\smallskip
  
\aut B. Elattari, J. Richert, P. Wagner and Y.M. Zheng*
\editeur Nucl. Phys.* {\bf A592} (1995) 385
\smallskip
  
\aut J.B. Elliott et al.*
\editeur Phys. Rev.* {\bf C49} (1994) 3185
\smallskip
  
\aut M.L. Gilkes et al.*
\editeur Phys. Rev. Lett.* {\bf 73} (1994) 1590
\smallskip
  
\aut J.B. Elliott et al.*
\editeur Phys. Rev.* {\bf C55} (1997) 1319
\suitaut B. Elattari, J. Richert, P. Wagner and A. Nourredine*
\livre Preprint CRN Strasbourg 95-37*, 1995, unpublished
\smallskip
  
\aut J.B. Elliott et al.*
\editeur Phys. Rev.* {\bf C55} (1997) 544
\smallskip
   
\aut W. Bauer and A. Botvina*
\editeur Phys. Rev.* {\bf C55} (1997) 546 and refs. therein
\smallskip
  
\aut Jac-Kwon Kim, Adauto J.F. de Souza and D.P. Landau*
\editeur Phys. Rev.* {\bf E54} (1996) 2291
\smallskip
  
\aut M.E. Fisher* in {\it Critical Phenomena}, Proceedings of the 51st 
Enrico Fermi School, edited by M.S. Green (Academic Press, New York, 1972)
\smallskip
  
\aut {\rm See f.i.}~ A. Aharony and D. Stauffer*
\livre Introduction to Percolation Theo\-ry, Taylor and Francis eds., 1994*
\smallskip
  
\aut J. Adler, Y. Meir, A. Aharony and A.B. Harris*
\editeur Phys. Rev.* {\bf B41} (1990) 9183
\smallskip
    
\aut J.M. Debierre*
\editeur Phys. Rev. Lett.* {\bf 78} (1997) 3145
\smallskip
  
\aut J. Richert, D. Boos\'e, A. Lejeune and P. Wagner*
\livre Contribution to the XXXV International Winter Meeting on Nuclear Physics,
Bormio, January 1997*
\smallskip
  
\aut X. Campi*
\editeur Phys. Lett.* {\bf B208} (1988) 351    
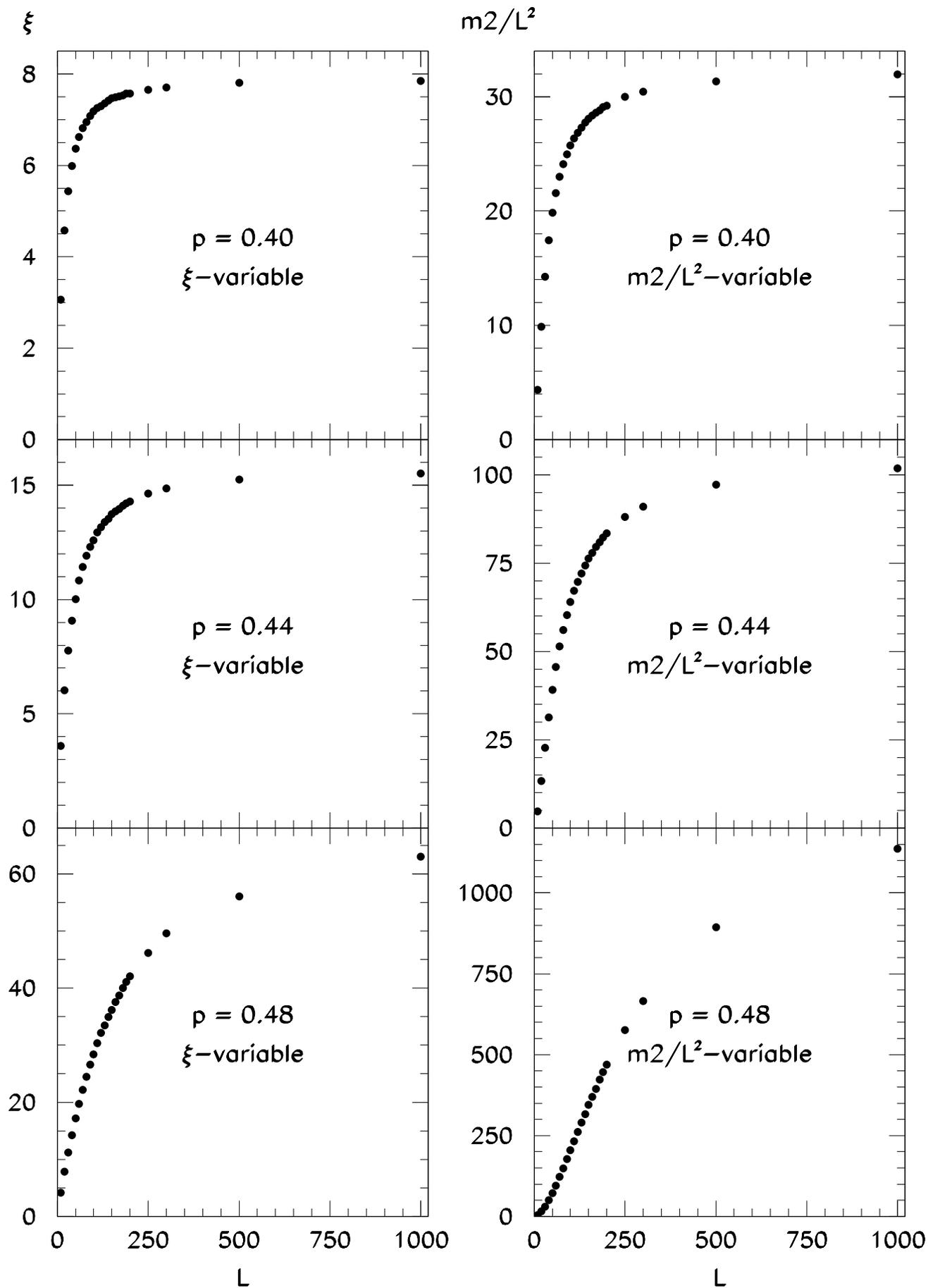
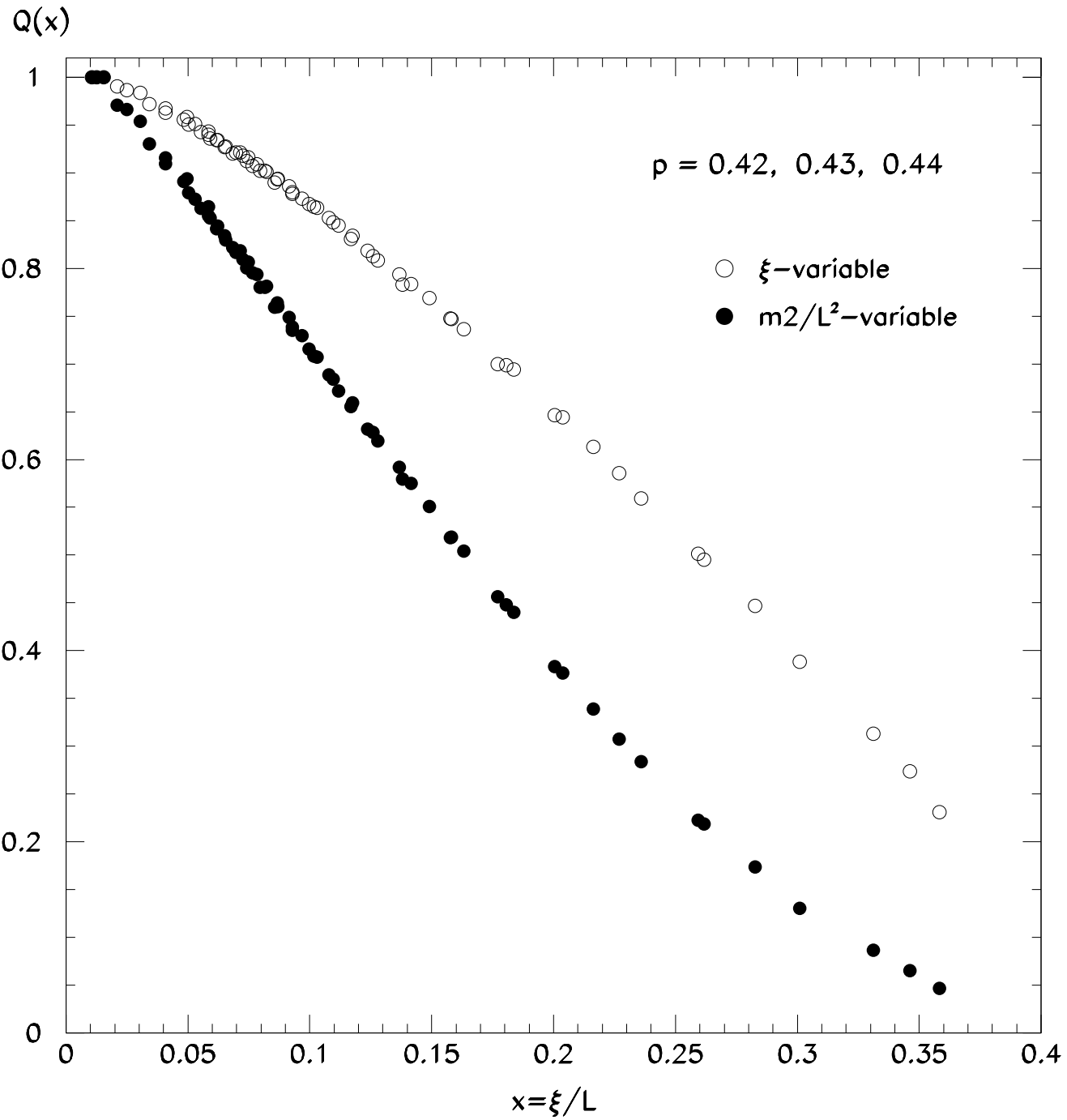
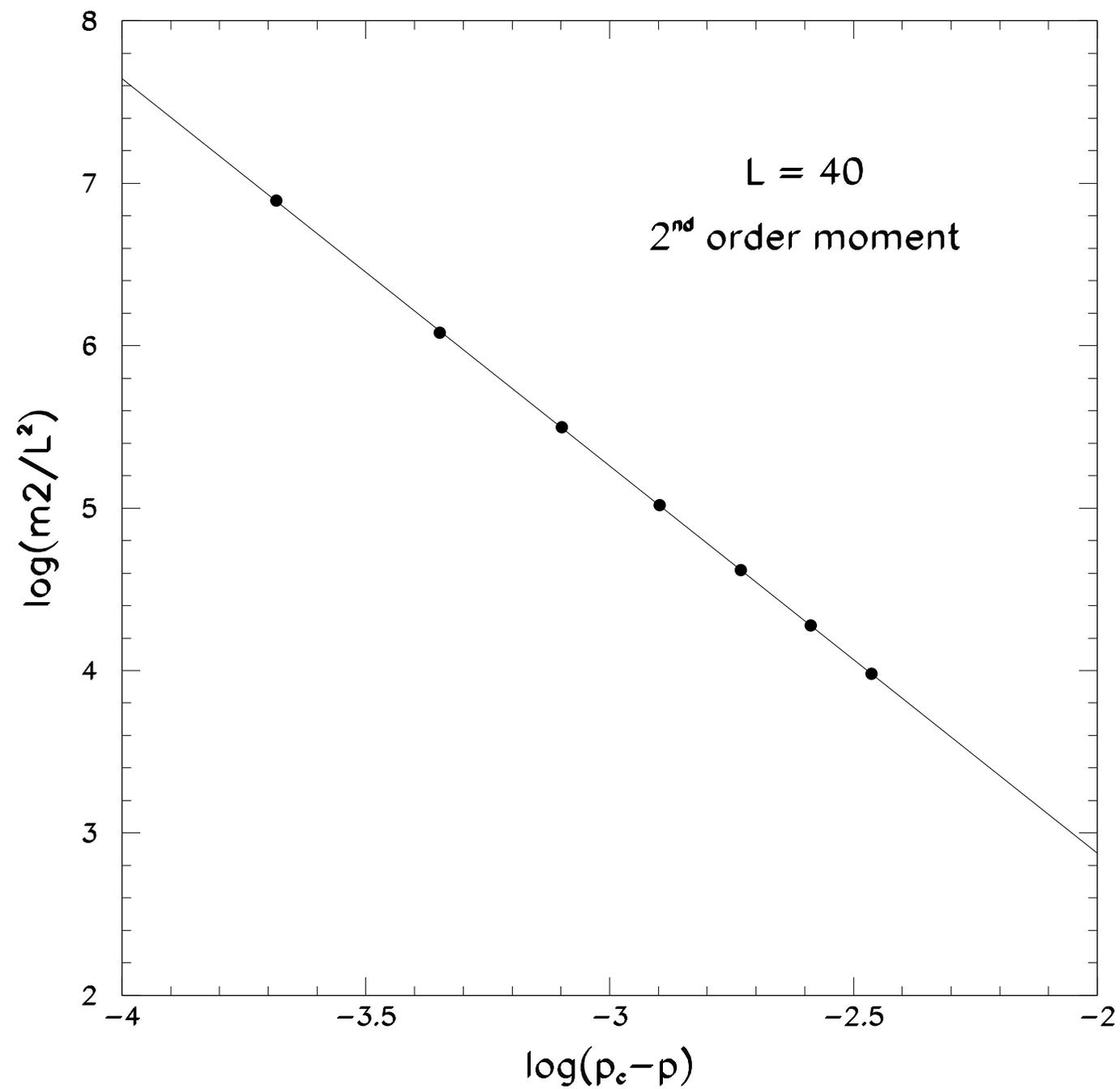
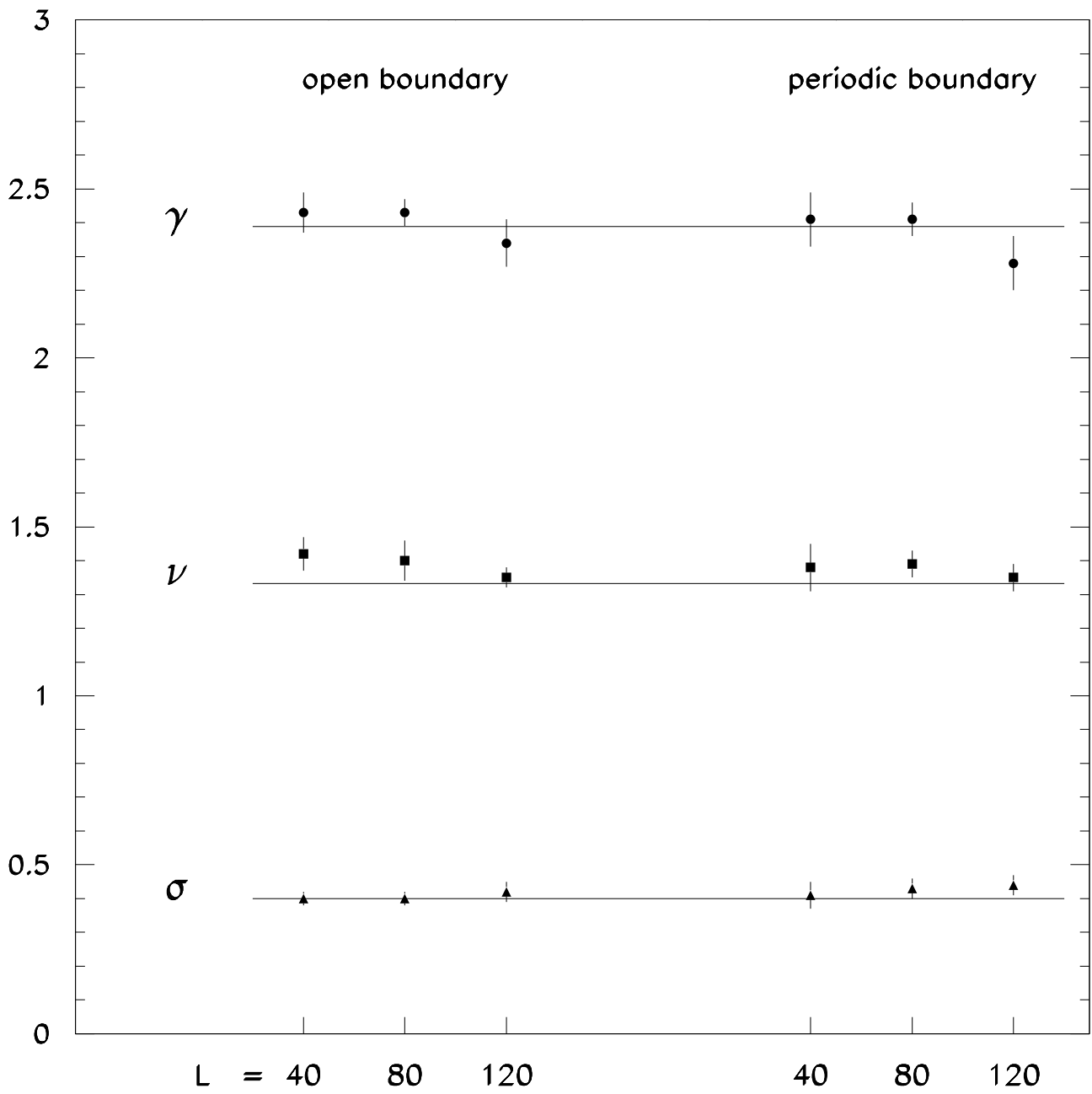
\bye